# Essential properties of the Difference Method for the Search of the Anisotropy of the Primary Cosmic Radiation

V.P. Pavlyuchenko[a)], R.M. Martirosov[b)], N.M. Nikolskaya[a)], A.D. Erlykin[a)]
*a) Lebedev Physical Institute, Moscow, Russia*
*b) Yerevan Physics Institute, Yerevan, Armenia*

*Corresponding author:* R.M. Martirosov,

(romenmartirosov@rambler.ru)

**Abstract.**
The methodical properties of the original difference method for the search of the anisotropy at the knee region of the primary cosmic radiation energy spectrum are analyzed. The main feature of the suggested method is a study of the difference in the EAS characteristics in different directions but not their intensity. It is shown that the method is stable to the random experimental errors and allows to separate the anomalies related to the laboratory coordinate system from the anomalies in the celestial coordinates. The method uses multiple scattering of the charge particles in the Galaxy magnetic fields to study the whole celestial sphere including the regions outside of the line of sight of the installation.

## Introduction

The charge particle energy spectrum of the primary cosmic radiation (PCR) is well described by the power law with an index of 2.7 from the energy $\sim 10^{10}$ eV up to the energy of $\sim 3 \times 10^{15}$ eV. At the higher energy the index increases rapidly to 3.1 creating the knee in the energy spectrum [1]. The main astrophysical hypotheses of the nature of the knee [2 - 4] related to the PCR origin, however, still do not have reliable experimental confirmation. It is explained by the confusion of the charged particles trajectories in the Galactic magnetic fields. As a result of this process the isotropy of the PCR at energies $10^{14} - 10^{16}$ eV is fulfilled with the accuracy not worse than 1% [5]. The propagation of the PCR particles is close to a Brownian motion and can be considered as a diffusive transport. Therefore, the main task now is to detect statistically valid anomalies of the PCR on the celestial sphere.

## Difference method for the analysis of experimental data [6,7]

This method initially takes into account the diffusive character of propagation of the PCR in the Galaxy. The method was developed specifically for the search of the anomalies and has a high sensitivity. It analyses the anomalies of the mass composition of the PCR and not the intensity of the particle fluxes from different parts of the sky. It is well known that at the equal conditions nuclei with different mass are scattered differently by magnetic fields. At the same time multiple scattering of particles in the Galaxy (diffusion) does not prevent and even helps the registration of the PCR from the sources located outside of the line of sight of the experimental installation.

According to the extensive air shower (EAS) experimental data the mass composition is estimated with the large errors by the indirect methods using the model dependent calculations. However, for the primary search of the anomalies it is necessary simply to determine the difference of characteristics of the PCR coming from opposite directions but not their absolute values. In this case it is possible to use well measured parameter that depends on the mass composition and least depends on the EAS development models.

The essence of the method is the following: the arbitrary directions $(l_0, b_0)$ are taken in the celestial coordinates. Then around each of them the cone with $\psi_0$ angle is built to divide the statistics in two equal sets of the number of events. For both sets the distributions of the analyzed parameter are built in the same intervals of the parameter. These distributions are subtracted from each other with the calculation of $\chi^2 = \sum_i \left( \Delta_i / \sigma_i \right)^2$, where $\Delta_i = m_i - m_i^{anti}$ is a difference between distributions in the range $i$ with error $\sigma_i = \sqrt{m_i + m_i^{anti} + 1} = \sqrt{n_i + 1}$. Here $n_i$ is the total number of events for both sets in the range $i$. This value does not depend on the given angles $(l_0, b_0)$, which is of crucial importance for the comparison of the $\chi^2$ values between themselves while searching for the maximum value. The mass compositions of the coming PCR from the opposite directions are most different in the direction where $\chi^2$ is maximal. It occurs due the separation of the charged particles by rigidity in the chaotic Galaxy magnetic fields.

In the process of subtraction of the distributions the common background and the methodical uncertainties are automatically eliminated because they are identical for both sets when the operation of the experimental installation is stable.

Cosine of the plane angle $H$ between direction $(l_0, b_0)$ and the EAS with angles $(l, b)$ for the spherical coordinate system is $H = Cos\psi = Sinb_0 Sinb + Cosb_0 Cosb Cos(l - l_0)$. This is why the shower inside the cone is at $H > H_0 = Cos\psi_0$ and is at $H \leq H_o$ outside.

## Experimental results

For the present study $3.38 \cdot 10^6$ EAS with primary energy $2\times10^{14} \div 10^{17}$ eV registered with the GAMMA array (Mt. Aragats, Armenia, 3200 m a.s.l.) in 2011 – 2013 have been used. The axes of the selected EAS were located inside the circle with 60 m radius from the center of the array. As a working parameter, the EAS "lateral" age parameter S derived by fitting of the lateral distribution function (LDF) of the EAS electrons (registered with the surface e/γ detectors) by the Nishimura-Kamata-Greisen (NKG) approximation was applied. The lateral S parameter characterizes a steepness of the NKG and is correlated with the longitudinal age parameter of the shower development. The parameter S was chosen because of its weak dependence on the primary energy near the EAS maximum. The S is rather reliably defined parameter and is usually applied for calculation of the EAS size $N_e$ and the primary energy $E_o$ [8]. This parameter depends on the mass composition because the heavy nuclei interact at the higher altitude in the atmosphere compared with the protons. At the same time the fragmentation of the primary energy and the cascade development of the EAS for the heavy nuclei are faster than for the protons. Therefore EAS generated by the heavy nuclei have a wide LDF with a big value of S.

The "blind search" of the anomalies near the Galactic plane was carried out. The only maximum in the direction $l_0 = 97° \pm 3°$, $b_0 = 5° \pm 3°$ has been discovered with $\chi^2/J = 57.6$ at the number of degrees of freedom $J = 17$. The $\chi^2$ is independent of the sign (+ or −) of the value $\Delta_i = m_i - m_i^{anti}$. Additional analysis showed that the anomaly is located on the opposite side in the direction $l_0 = 277°$, $b_0 = -5°$ (equatorial coordinates α =140°, δ = -57°, Southern Celestial Hemisphere) and is related to the knee region of the primary energy spectrum. The more detailed description of the GAMMA array and the results derived from the installation were published elsewhere [6, 7, 9, 10].

## Analysis of method

Firstly the value $\chi^2/J = 57.6$ in the maximum at the 17 degrees of freedom has been verified. It is a very large value. At the random dispersion this value must be close to $1 \pm \sqrt{2/J}$ and each unit increases associated statistical significance of the deviation of the result from a random fluctuation by about one σ. Hence, the significance of derived result is more than 50 σ. With the division of the total number of events by 2, 3, 5 and 10 we are convinced that this value is that large mainly because of the large statistical data and the high sensitivity of the method which uses the full statistics for each point in the sky. In order to decrease the statistical sample but at the same time to least change conditions of the EAS registration every second, third, fifth and tenth events were selected for the analysis. The maximum value of $\chi^2/J$ decreased from 57.6 to 29.4, 20.3, 12.6 and 7.9, respectively. It means that in our case the value $(\chi^2/J)-1$ linearly depends on statistical sample with the practically constant shape of the distributions.

$\chi^2/J$ can also be systematically overestimated due to the number of reasons. In order to verify it the direction where $\chi^2/J$ is minimal has been found. This minimum is equal to 1.32 at $l_0 = 15°$, $b_0 = 60°$ and within of standard deviation coincides with the random distribution of $\Delta/\sigma$. Both verification tests described above confirm high statistical reliability of the anomaly.

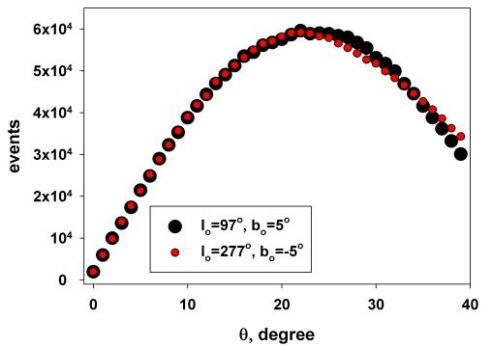 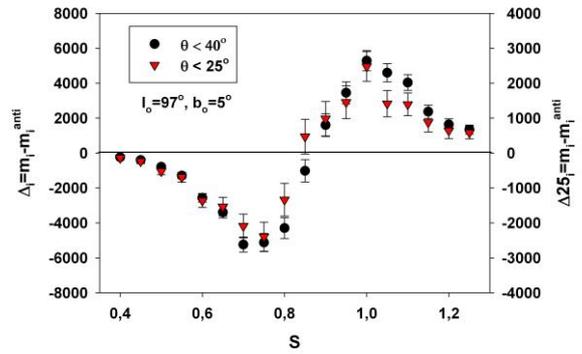

*Fig. 1-1. EAS distribution of zenith angle $\theta$ at maximum of $\chi^2/J$ for opposite directions*

*Fig. 1-2. Dependence of differences $\Delta_i = m_i - m_i^{anti}$ on S parameter for all EAS (left scale) and for EAS with $\theta < 25°$ (right scale)*

The asymmetry of the installation by the zenith or azimuthal angles can also imitate the anomaly. The range of the EAS zenith angles which were used for the analysis is $0° \div 40°$. The EAS development and registration conditions for $\theta = 40°$ are different from the vertical EAS. We derived the EAS distributions of zenith angles $\theta$ for opposite directions in the region of anomaly (*Fig. 1-1*). It can be seen that the distributions are completely the same up to $\theta < 25°$. There are some differences at $\theta > 25°$. The set of $1.96 \times 10^6$ showers with $\theta < 25°$ have been examined separately. The coordinates of the maximum as well as the shape of the distribution of $\Delta_i = m_i - m_i^{anti}$ in the maximum remained within the error limits (*Fig. 1-2*). The value of the maximum decreased from 57.6 to 19.1 because of the decrease of the statistical sample and the narrowing of the EAS selection sector. It is observed that differences are negative for the small values of $S$, i.e. the "young" EAS are coming from the opposite direction (from the Southern hemisphere).

The anomaly could be imitated by the geomagnetic field, because the electrons of the EAS developing along the magnetic field will also tend to move along the field lines. It can lead to the narrowing of the LDF and to the decreasing of S compared with the EAS coming from other directions. The zenith angle of the geomagnetic field at the region of the GAMMA array is about 25º (magnetic inclination is ≈ 65°). Therefore, the largest influence of the geomagnetic field on the zenith angle dependence of S parameter should be in the southern direction at $1/\cos \theta \approx 1.1$ (*Fig. 2-1*). The figure demonstrates the linear dependence of S parameter in the direction 97°, 5° and in opposite direction from value $1/\cos\theta$, which is proportional to the thickness of atmosphere above the installation. No significant difference from the linear dependence was found in the whole range of the zenith angles. The figure also shows directly that a "young" EAS are coming from the southern direction at any values of θ. The similar properties and linear dependences from S are observed in the other angles of the celestial sphere. The differences between the values of S parameters for direct and opposite directions are changing slightly with the change of the celestial coordinates.

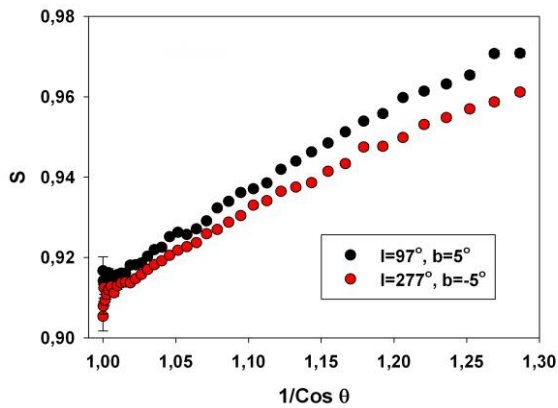 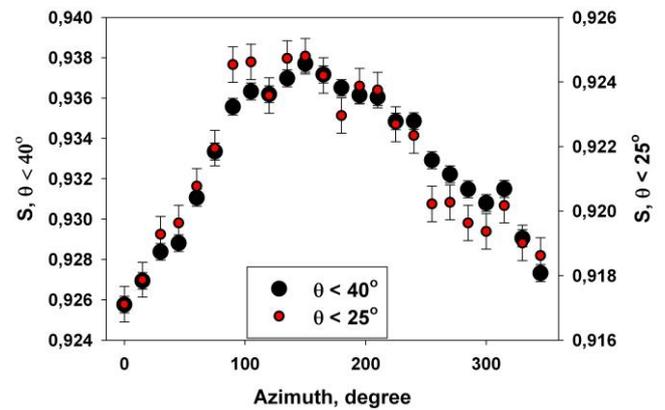

*Fig. 2-1. Dependence of parameter S from zenith angle $\theta$*

*Fig. 2-2. Dependence of parameter S from azimuthal angle for EAS with $\theta < 40°$ and $\theta < 25°$*

It can be concluded that the coordinates and the properties of the anomaly are stable with respect to the range of the EAS zenith angles, but narrowing of the selection range decreases the reliability of the result.

To test the influence of asymmetry of the azimuth angles on the anomaly, the dependence of the parameter *S* from the azimuth angle A in the horizontal astronomical (laboratory) coordinate system (where "0°" is directed to the South) has been obtained (*Fig. 2-2*). The small regular and almost sinusoidal dependence of S from A is observed with identical shape for the both ranges of zenith angles $\theta$.

This dependence can be both the cause of the anomaly and its consequence. Let's assume that dependence S(A) has arisen because of some asymmetry of the experimental installation and this is the cause of the anomaly. In this case application of correction, eliminating the dependence S(A), should lead to the change of coordinates or to the destruction of the anomaly. To verify that the corresponding corrections $\delta(A) = S(A) - <S>$ were subtracted from the S for all selected EAS. Then, as before, the anomaly has been searched (*Fig. 3-1*).

It is seen from *Fig. 3-1* that the position of the main maximum did not change within the experimental errors but the value $\chi^2/J$ in the maximum decreased to 26.3. In addition, the axially symmetric second maximum $\chi^2/J = 20.3$ has appeared with coordinates $l_0 = 123°$, $b_0 = 27°$ which coincides with the direction of the Earth's rotation axis – the Celestial Pole.

The decrease in the main maximum and appearance of an additional maximum at the Pole mean that the application of the correction doesn't improve the initial experimental data but made them even worse imitating the asymmetry of the experimental installation. Indeed, any asymmetry of the installation in a coordinate system rotating with the Earth can be divided into two components: parallel and perpendicular to the Earth rotation axis. During the three years of operation of the installation the Earth has made more than one thousand rotations. Therefore, the component perpendicular to the axis of the installation was annihilated with a good precision as $\int_0^{2\pi} SinAdA$. This analysis points out to the coincidence of the maximum and the position of the Pole with the 1° accuracy. At the same time, the parallel component was summed up and appeared in the form of the peak.

This statement was also verified by using the another approach: the value of S was increased by 0.07 for all EAS with azimuth A = 45° ÷ 60° imitating the anisotropy in this direction. The value of 0.07 was chosen to have the noticeable, but not catastrophic effect. The result is shown in *Fig. 3-2*. It demonstrates once again that the main peak did not change its position but decreased to 34.5, and again the second peak appeared at the Pole with the maximum at 34.6 due to the artificially created anisotropy. *Fig. 3-3* shows the result when the value S was increased by 0.1 for all EAS with the azimuth of A = 300° ÷ 315°, perpendicular to the previous one. The effect has remained qualitatively the same: the main maximum has decreased to 33.4, but maximum on the Pole has increased to 88.2 (its top has been cut to 50 on the figure to insure the visibility of the main peak).

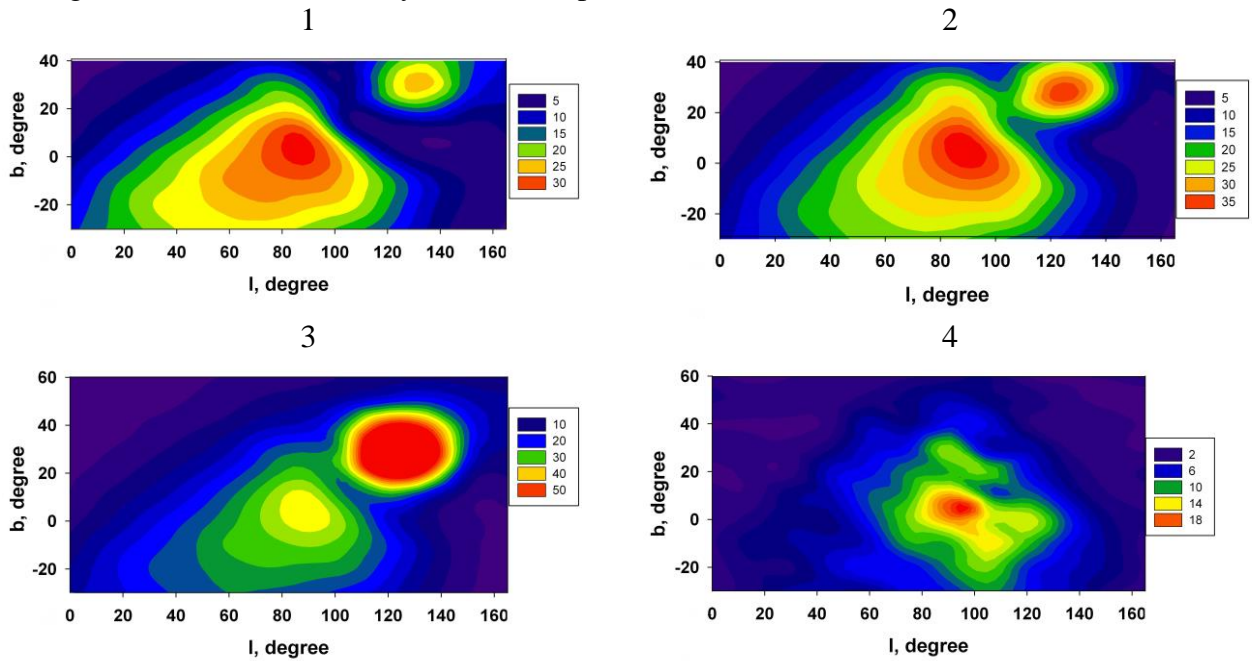

*Fig. 3. Dependence of $\chi^2/J$ from the Galactic coordinates (l, b). Contour diagrams.*
*1 – corrected by the alignment of dependence of S from azimuth;*
*2 – value S was increased by 0.07 for the range A = 45° – 60°.*
*3 – value S was increased by 0.1 for the range A = 300° – 315°.*
*4 – value S was increased by random number simulated by Gaussian with the average values equal to 0 and $\sigma$ = 0.25*

In order to verify that the peak occurring at the Pole is due to the asymmetry in the laboratory coordinate system, but not just due to the random errors in the calculations of S, a random number, which was simulated by Gauss distribution with the mean "0" and with different $\sigma$, was added to the initial value S in all events independently of azimuth. The main maximum doesn't change its position but decreases to 42.7 at $\sigma$ = 0.1 and to 19.4 at $\sigma$ = 0.25 (*Fig. 3-4*). Considering that the initial distribution of the S parameter has $\sigma$ = 0.16, the stability of the position of the maximum for such big distortions shows a stability of the method relative to the random experimental errors.

Returning back to *Fig. 2-2* it can be concluded that discovered dependence of S from azimuth is related to the celestial coordinates and indicates the arrival of "younger" EAS (more light primary particles) from the Southern hemisphere.

The solar-daily geophysical changes can also have an effect on the obtained coordinates. The dependence of S from the time of the day is presented on *Fig. 4-1*, and shows a weak regular dependence. After inclusion of a corresponding correction the maximum slightly increased from

57.6 to 58.7. Thus, the change is related to the Earth coordinate system and introduces a negligible error. The value $\chi^2/J$ on the Pole did not change.

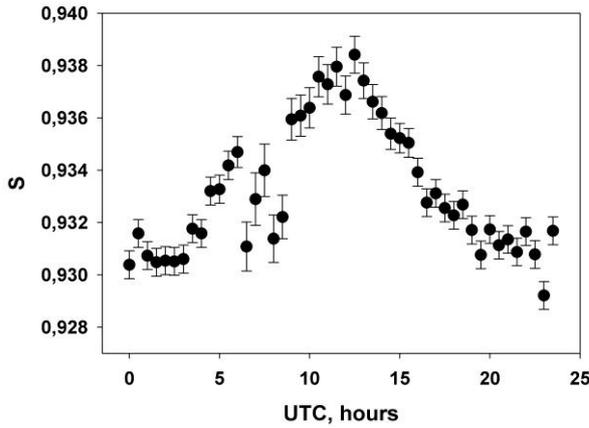 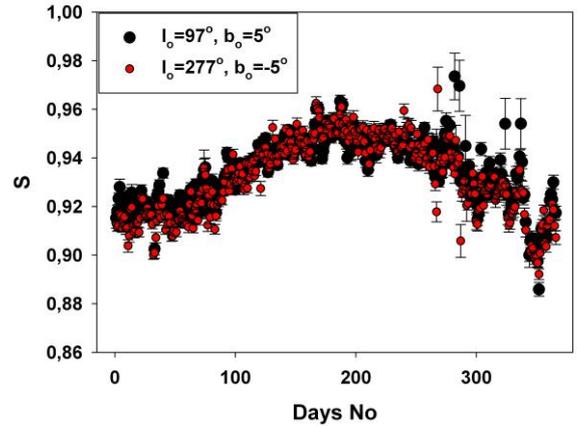

Fig. 4-1. Dependence of S from the time of the day in UTC. Between 5 – 11 hr (10 -16 local time) the adjustment of the installation was performed.

Fig. 4-2. Dependence of S from the day of the year for EAS set in the direction 97°, 5° and in opposite direction 277°, -5°

Regular dependence of S from the day of the year (*Fig. 4-2*) with the amplitude of almost an order of magnitude more than for the sun-daily dependence was found. Its shape coincides with the seasonal behavior of the temperature that might be an evidence of rather strong dependence of S parameter from the air temperature. In the *Fig. 4-2* S distributions for the sets of events from the opposite directions coincide well. Thus, when the sets are subtracted to estimate the value of $\chi^2/J$, dependence of S parameter from the day of the year should be annihilated or strongly reduced. As mentioned above, the background and methodical uncertainties are automatically annihilated as they are identical for both sets. It should also be noted that "younger" EAS are coming from the place of the anomaly (day of the year is ≈220) in comparison with the showers from opposite direction (*Fig. 1-2, 2-1*), and they are in antiphase from the temperature dependence.

For all verifications the same statistical sample size ($3.38 \times 10^6$ events) has been used.

Following conclusion can be drawn from all the described verifications: random errors of the S parameter do not change the position of the anomaly, however, they reduce the maximum value of $\chi^2/J$, which depends on the magnitude of distortion. If distortions are also azimuthally asymmetric in the system, and rotate together with the Earth, the maximum of $\chi^2/J$ appears in the direction of the Earth's rotation axis.

Consequently, the difference method not only suppresses the background and methodological uncertainties, but also allows to separate the effects associated with the laboratory coordinate system from the effects related to the fixed system of celestial coordinates. Detection of the peak on the Pole illustrates the azimuthal asymmetry of the experimental setup, whereas, its absence indicates the lack of asymmetry or an effective compensation of such.

## Discussion of results

Investigation of the properties of the difference method using a quantitative criterion $\chi^2/J$ for the EAS age parameter S has shown the high sensitivity and stability of the method. The method has a unique property that allows separation of the effects associated with the laboratory coordinate system from the effects related to the fixed system of celestial coordinates. The peak at the Pole points out to the anomalies associated with the solar - daily rotation of the Earth. Sustainable anomaly in other directions is likely related to the celestial coordinates or to the annual cycle.

Near the discovered anomaly (statistical reliability > 50 σ) with the Galactic coordinates $l_0 = 277°$, $b_0 = -5°$ there is a cluster in the constellation Vela with two closely located supernova remnants Vela X (263.9°, -3.3°) and Vela Jr (266.2°, -1.2°), at the approximate distances from Earth of 0.3 and 0.2 kpc, respectively. It is obvious that this cluster is a good candidate for a nearby source of the PCR associated with the formation of the knee. The cluster is close to the direction of the detected anomaly, it is located not far from the Earth and its energy is close to the energy of the knee. Unfortunately, the difference method does not allow to determine the intensity of the particle flux coming from the source.

The excess of "young" EAS coming from this direction is most likely related to the mechanism of diffusion along the path from the source to the Earth. The younger showers characterize the lighter mass composition of the PCR with a predominance of the protons and helium nuclei. In diffuse transport the heavier nuclei are more deflected in the magnetic fields. So the flux of the PCR from the source to the Earth is depleted by the heavy particles leading to the lighter mass composition with the "younger" showers in comparison with the EAS from the opposite side.

Relatively strong change of the "lateral" age parameter S dependent on the day of the year has been found, coinciding with the seasonal behavior of the air temperature. This effect requires verification by direct measurement of the dependence of S from the temperature. And if it is really significant, it should be taken into account in the analysis of experimental data and simulating of the cascade processes in the atmosphere. The age parameter S is one of the basic characteristics of EAS, since it is used for calculation of the shower size $N_e$, primary energy $E_o$ and mass composition of the PCR. Unfortunately, the GAMMA installation, which has now completed its work, did not register the data on the condition of the atmosphere.

**Conclusion**

The difference method has shown its simplicity, high sensitivity and stability to the random errors. The method is able to study the full celestial sphere with the facility with limited field of view in the laboratory coordinate system, as well as to separate anomalies in the system of coordinates related with the daily rotation of the Earth from the anomalies in the celestial coordinates. The observed anomaly is a good candidate for the nearby source of the PCR associated with the formation of the knee.

The main difference from the traditional methods is a study the EAS characteristics in the different directions, and not their intensity. Together with the age parameter S other EAS characteristics and their combinations can be used as the experimental parameters.

**Acknowledgments**

We are grateful to all colleagues at the Moscow Lebedev Physical Institute and the Yerevan Physics Institute who took part in the development and exploitation of the GAMMA array. We are also grateful to the administration of the Department of Nuclear Physics and Astrophysics of the Moscow Lebedev Physical Institute, to the Yerevan Physics Institute, to DESY as well as to the State Committee of Science of the Republic of Armenia, to the ''Hayastan'' All-Armenian Fund and to Program of Fundamental Research of the Presidium of the Russian Academy of Science "Fundamental properties of matter and astrophysics" for financial support.